# Project Lyra: Catching 1I/'Oumuamua Using Nuclear Thermal Rockets

Adam Hibberd, Andreas M. Hein

adam.hibberd@ntlworld.com, andreas.hein@i4is.org

Initiative for Interstellar Studies (i4is)

27/29 South Lambeth Road London, SW8 1SZ United Kingdom

## Abstract
The first definite interstellar object observed in our solar system was discovered in October of 2017 and was subsequently designated 1I/'Oumuamua. In addition to its extrasolar origin, observations and analysis of this object indicate some unusual features which can only be explained by in-situ exploration. For this purpose, various spacecraft intercept missions have been proposed. Their propulsion schemes have been chemical, exploiting a Jupiter and Solar Oberth Maneuver (mission duration of 22 years) and also using Earth-based lasers to propel laser sails (1-2 years), both with launch dates in 2030. For the former, mission durations are quite prolonged and for the latter, the necessary laser infrastructure may not be in place by 2030. In this study Nuclear Thermal Propulsion (NTP) is examined which has yet to materialise as far as real missions are concerned, but due to its research and development in the NASA Rover/NERVA programs, actually has a higher TRL than laser propulsion. Various solid reactor core options are studied, using either engines directly derived from the NASA programs, or more advanced options, like a proposed particle bed NTP system. With specific impulses at least twice those of chemical rockets, NTP opens the opportunity for much higher $\Delta V$ budgets, allowing simpler and more direct, time-saving trajectories to be exploited. For example a spacecraft with an upgraded NERVA/Pewee-class NTP travelling along an Earth-Jupiter-1I trajectory, would reach 1I/'Oumuamua within 14 years of a launch in 2031. The payload mass to 1I/'Oumuamua would be around 2.5metric tonnes, but even larger masses and shorter mission durations can be achieved with some of the more advanced NTP options studied. In all 4 different proposed NTP systems and 5 different trajectory scenarios are examined.

## 1. Introduction
Now designated 1I/'Oumuamua, the first definite interstellar object was discovered within our solar system in October 2017 by the PanSTARRS 1 observatory [1-3]. Its discovery and subsequent observation sparked significant interest in both academia and the media. Possessing various unusual characteristics, debate has ranged from 1I/'Oumuamua's shape [4-9], its composition [4,5,11-15], its origin [10,16-25], the possible abundance of such objects [2, 5] and also the nature of a non-gravitational perturbing acceleration [26,27]. A second interstellar object, 2I/Borisov was discovered in August 2019 [17,18] but apart from certain unusual properties [19,20], does not seem to present the enigma that is 1I/'Oumuamua and so 1I/'Oumuamua possibly remains the more compelling option for a spacecraft mission. Indeed an in-situ study of the object would be of considerable scientific value. A spacecraft taking a direct trajectory from Earth has been proposed by Seligman and Laughlin [28], but is impracticable for 1I/'Oumuamua because the optimum launch window was before its discovery.

Hein et al [29,30] have shown that missions to 1I would be feasible using existing and near-future chemical rocket technology combined with a trajectory to Jupiter, followed by a Solar Oberth maneuver. Further study by Hibberd et al [31] revealed that, preceded by a 'V$_\infty$ Leveraging Maneuver' and employing a Solar Oberth at 6 Solar Radii, a trajectory can achieve lower ΔV's and also possess a periodicity in launch optima of approximately 12 years. These missions have assumed chemical propulsion.

Nuclear Thermal Rockets (NTR) have the advantage of significantly higher specific impulse compared to chemical (a factor > 2), enabling lower interplanetary flight times or equivalently achieving higher ΔV's. Furthermore as a consequence of the NASA Rover/NERVA research and development programs from 1955-1972 [42], they also have a high TRL (5-6). They are recognized by NASA as a possible game-changer as far as manned missions to Mars are concerned, and are the default propulsion system in the Mars Design Reference Architecture (DRA) 5.0 study [43]. Nevertheless, the existing literature has not yet explored the potential of NTR to achieve intercept of Interstellar Objects. Despite its relatively high maturity, NTR would still require further maturation before it could be operated in space. Nevertheless, flight demonstration is expected for the 2020s.

Here the focus is on the ISO 1I/'Oumuamua. For 1I/'Oumuamua, its discovery occurred after perihelion and was after the optimum launch date for intercept. Hence, unless a Comet-Intercepor type architecure can be exploited (which will be launched in 2028 and will not able to reach 1I/'Oumuamua), an ISO may be receding from the sun at speeds of tens of km/s by the time a mission can be mounted, and probably at a high orbital inclination, imposing a considerable challenge as far as an intercept mission is concerned. Can NTR meet this challenge ? The following paper addresses this question by firstly examining some NTR options and their ΔV capability. Different trajectory options are then proposed. The performance, in terms of minimizing flight duration of these trajectories is derived using Optimum Interplanetary Trajectory Software (OITS), and then related to spacecraft payload mass, based on the different NTR options.

## 2. Materials and Methods
### 2.1 Optimum Interplanetary Trajectory Software

In order to compute the interplanetary trajectories to 1I and their associated ΔVs (high impulsive thrust is assumed), the Optimum Interplanetary Trajectory Software (OITS) [35] developed by Adam Hibberd is utilized. It adopts the patched conic assumption, where the s/c is either gravitationally attracted to the Sun (which will be the case for most of the trajectory) or to a celestial body, if the s/c lies within that body's sphere of influence. For calculating position and velocity as a function of time, OITS exploits the NASA SPICE toolkit and corresponding binary SPICE kernel files are used.

For OITS, the user selects a sequence of celestial bodies for the s/c to visit, starting with Earth and ending with the target body, in this case 1I/'Oumuamua. For OITS, the independent variables which completely specify an interplanetary trajectory, and which are optimized by the software, are the times of encounter of the s/c at each celestial body in turn. The software then seeks to find trajectories which, depending on the version of OITS, either minimize ΔV, or minimize overall mission duration.To do this, OITS first solves the Lambert problem between each adjacent pair of celestial bodies using the Universal Variable Formulation [32].Second OITS computes the encounter ΔV at each of the celestial bodies, assuming this ΔV is applied at the periapsis point w.r.t. the celestial body, in a direction perpendicular to the radial vector, and in the plane defined by the incoming approach velocity of the s/c and the departure velocity of the s/c. These two velocities can be computed from the interplanetary trajectories either side of the celestial body in question. As an alternative, instead of a celestial body, the user can select an 'Intermediate Point'(IP) which is a massless and static point sitting on a sphere centred at the origin of the Solar System and with a user-specified radius. Thus the heliocentric longitude and latitude polar angles of this IP can then be optimized by the software, in addition to the encounter times. All this can be expressed mathematically as a non-linear global optimization problem with inequality constraints and is solved by applying the NOMAD solver [33]. A definition manual detailing the theory behind OITS can be found on github [35].

The previously mentioned assumptions will clearly introduce inaccuracies and the solutions will not necessarily be precisely optimum as can be found in [36], however solutions generally lie within 1% ΔV of NASA's trajectory browser.

For all planetary encounters a minimum limit periapsis altitude of 200km is specified relative to the planet's equatorial radius. In the case of Jupiter, this equatorial radius is taken at the 1bar level, i.e. 71492km.

## 2.2 NTR Propulsion Options

Four NTR options are considered here and are provided in Table 1. These NTR engines are also listed in [37].

*Table 1 : NTR options with their performance values*

| NTR motor | Description | Ref. | Mass/kg | Specific Impusle, Isp (s) | Thrust (kN) |
|---|---|---|---|---|---|
| NERVA/Pewee Class | Upgraded version of the Pewee studied in the NASA NTR NERVA program '50s to early '70s | [38] | 3250 | 906 | 111.7 |
| SNRE | Based on the Small Nuclear Rocket Engine, studied in the Rover program | [38] | 2400 | 900 | 73 |
| SLHC | Square Lattice Honeycomb | [39] | 2500 | 970 | 147.5 |
| SNTP | Particle Bed Nuclear Thermal Rocket | [40] | 800 | 950 | 196 |

It is assumed the s/c is transported to a LEO of 406km by a NASA Space Launch System (SLS) Block 2. It is currently envisaged that an SLS offers a 130metric tonne capability to LEO. We further assume that only one NTR motor is used, and $LH_2$ propellant can be stored for significant durations with no-leakage and with a zero boil-off cryocooler [41], and further that there are 2 Staged $LH_2$ tanks with an optimum mass ratio. Payload here is understood to mean the total mass of spacecraft after the engines and spent $LH_2$ tanks have been jettisoned. This therefore represents the useful available spacecraft mass. Note howeve\a\r that this mass includes that of any heat shield which may be necessary if a Solar Oberth is involved.

We get Figure (1) for payload mass against ΔV budget. The ratio of dry stage mass to wet stage mass for both stages is assumed to be p=0.035 which was calculated using the same spacecraft/$LH_2$ mass budgets provided in [41] and incorporates the tank insulation and cryocooler mass.

To generate Figure (2) for Ammonia ($NH_3$) propellant a factor of 0.63 is applied to the specific impulses provided for $LH_2$ above. The value of p for $LH_2$ is retained. This is probably an overestimate as the requirements on storage/insulation would be less severe than for $LH_2$, so this Figure (2) is probably a conservative estimate.

The data from Figures (1) & (2) will be used in Section (3.2) as input to generate flight times as a function of payload mass to 1I/'Oumuamua.

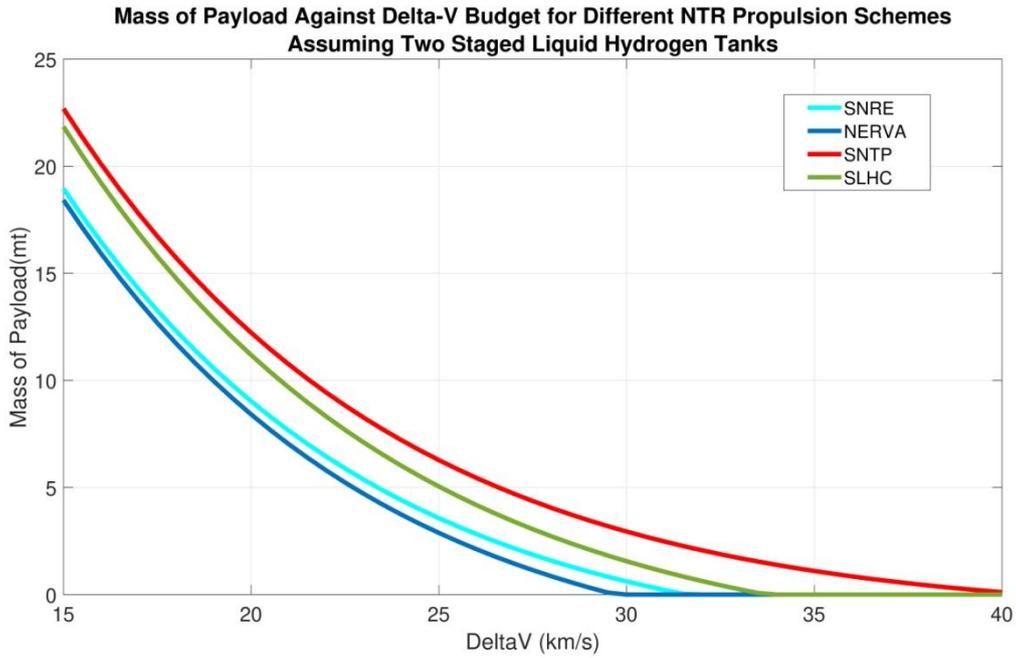

*Figure 1 : Mass of Payload against ΔV for four different NTR options and LH2 tanks*

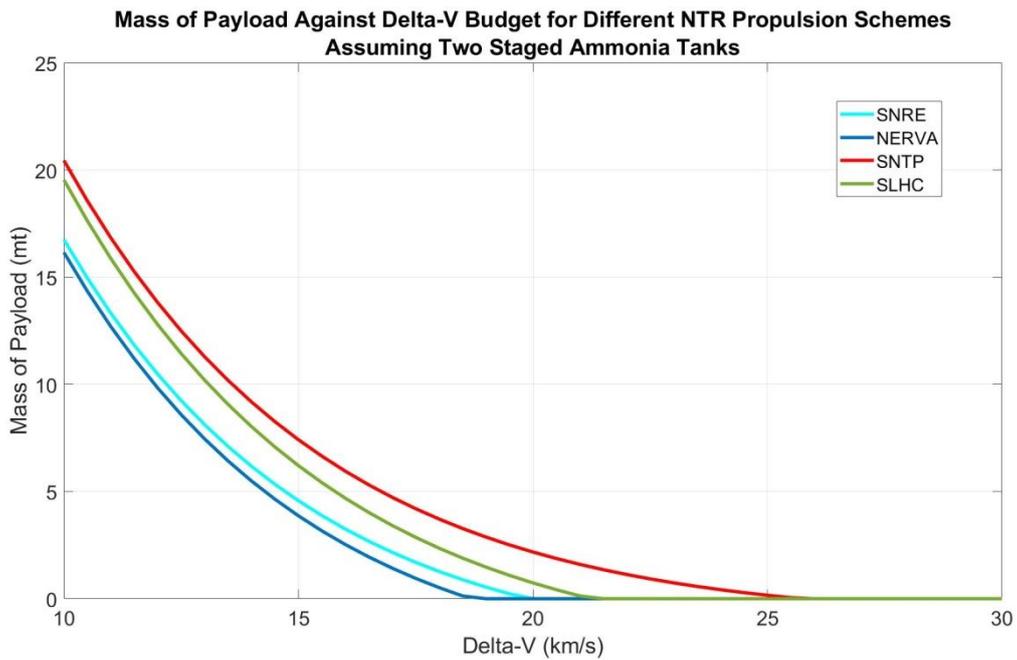

*Figure 2 : Mass of Payload against ΔV for four different NTR options and NH3 tanks*

## 2.3 Trajectory Scenarios

There are five scenarios considered here, which partly correspond to scenarios which have already been presented in the previous literature, but using chemical propulsion:

1) Direct from Earth to 1I/'Oumuamua[28-30]
2) From Earth to 1I using a Solar Oberth
3) From Earth to 1I via Jupiter and Solar Oberth[29-31]
4) From Earth to Jupiter to 1I using $LH_2$
5) Identical to (4) but using $NH_3$

Figures (3), (4), (5) show examples of trajectory senarios (2), (3), (4/5) respecitvely for illustration. Note that scenarios(1) & (2) have optimum trajectories every one Earth year, due to the position of the Earth with respect to 1I/'Oumuamua. Scenarios (3), (4) & (5) have optimum trajectories approximately every Jupiter year, so around 12 Earth years, due to the alignment of Jupiter with 'Oumuamua. Scenario (3) has optima in 2033, 2045, 2057 and so on. Scenarios (4) & (5) have launch optima in 2031, 2043, 2055 etc.

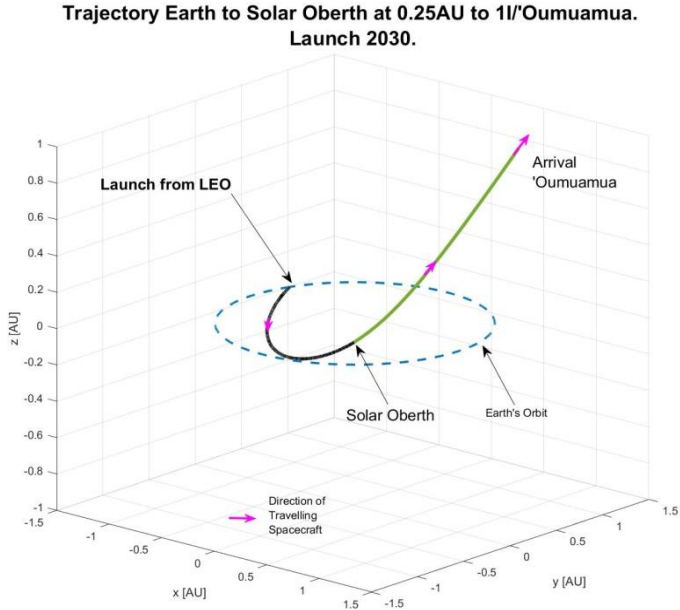

*Figure 3 : Example of Trajectory Scenario (2)*

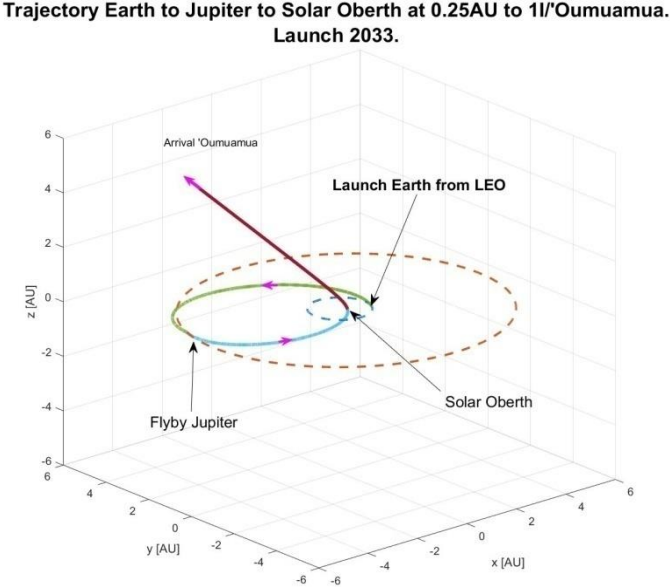

*Figure 4 : Example of Trajectory Scenario (3)*

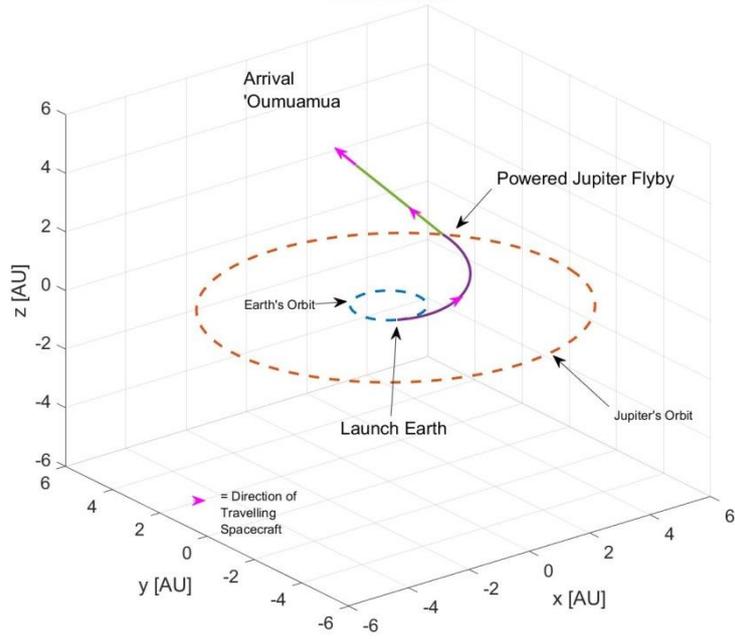

*Figure 5 : Example of Trajectory Scenarios (4) & (5)*

# 3. Results
## 3.1 Mission Flight Duration and ΔV Budgets for Different Trajectory Scenarios

The available ΔV for NTR is generally larger than that for chemical propulsion.

Figure (6) shows minimum flight duration againt launch date (based on yearly optima) for the years 2025 to 2045 and taking a direct trajectory (scenario (1) ). Four different ΔV budgets are allocated, 25km/s, 30km/s , 35km/s and 40km/s. The s/c is assumed to have already been placed in LEO of altitude 406km. Observe that for ΔV budgets of 25km/s and with launch years > 2035 , direct trajectories from LEO to 1I have prohibitively long mission durations.

Minimum flight duration plots against Solar Oberth perihelion solar radial distance are shown for three different ΔV budgets for scenario (2) in Figure (7) and four ΔV budgets for scenario (3) in Figure (8). Scenarios (4) & (5) minimum flight duration against ΔV is shown in Figure (9). For Figures (6), (7), (8) & (9) it is assumed the s/c starts in an LEO of altitude 406km.

Even with the powerful Space Launch System (SLS) Block 2, it can be shown that for scenarios (1) , (2) & (4/5), chemical propulsion cannot deliver ΔV's of the magnitude needed for sensible flight durations but scenario (3) is achievable.

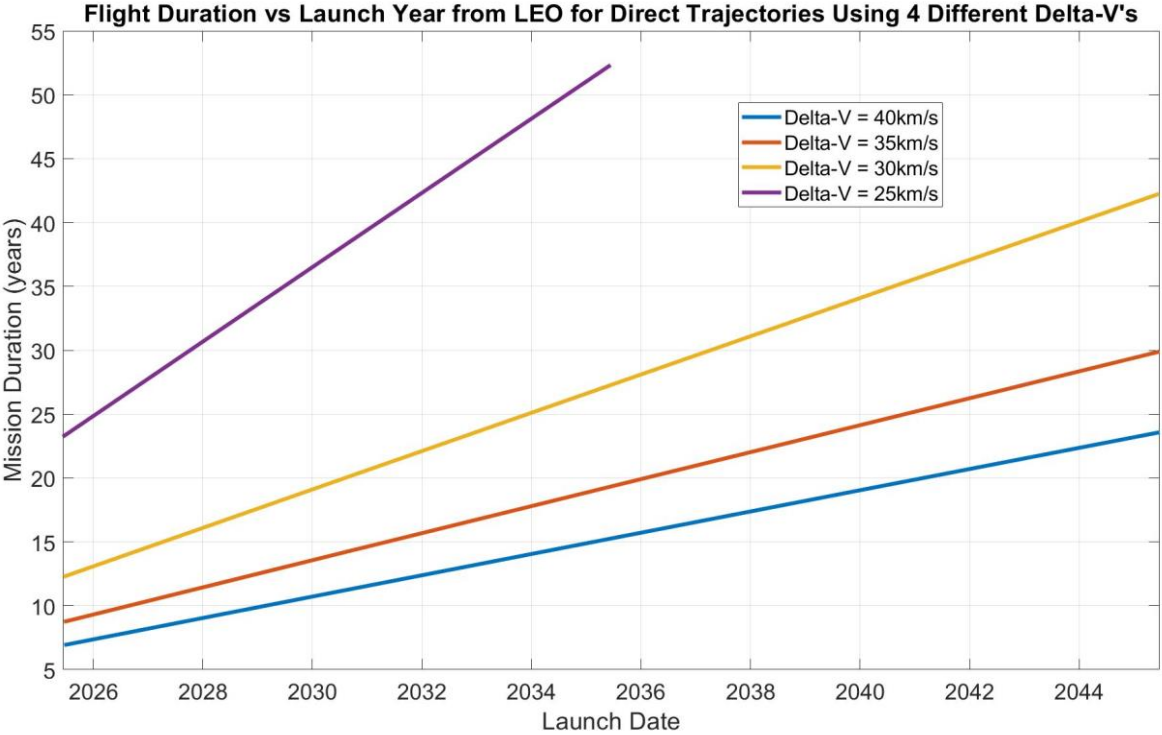

*Figure 6 : Scenario (1) Mission Duration vs Launch Date (assuming yearly optima launch dates)*

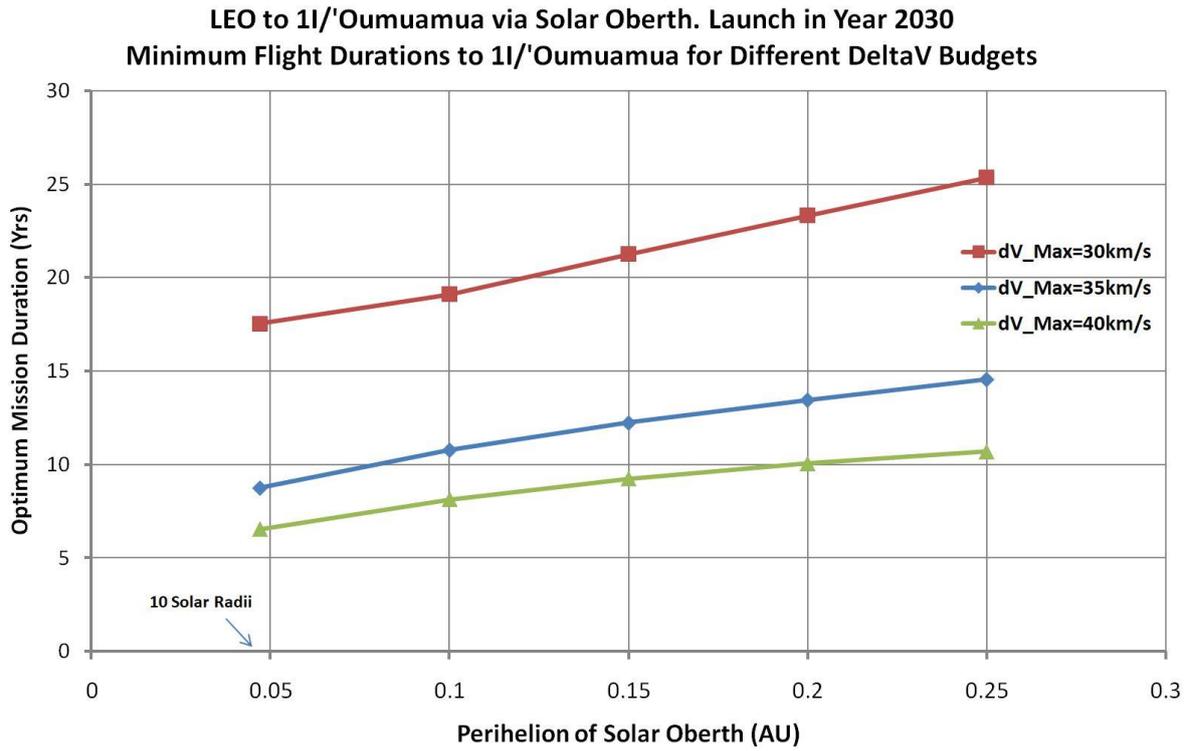

*Figure 7 : Scenario (2) Minimum Mission Durations vs Solar Oberth Distance for 3 ΔV budgets*

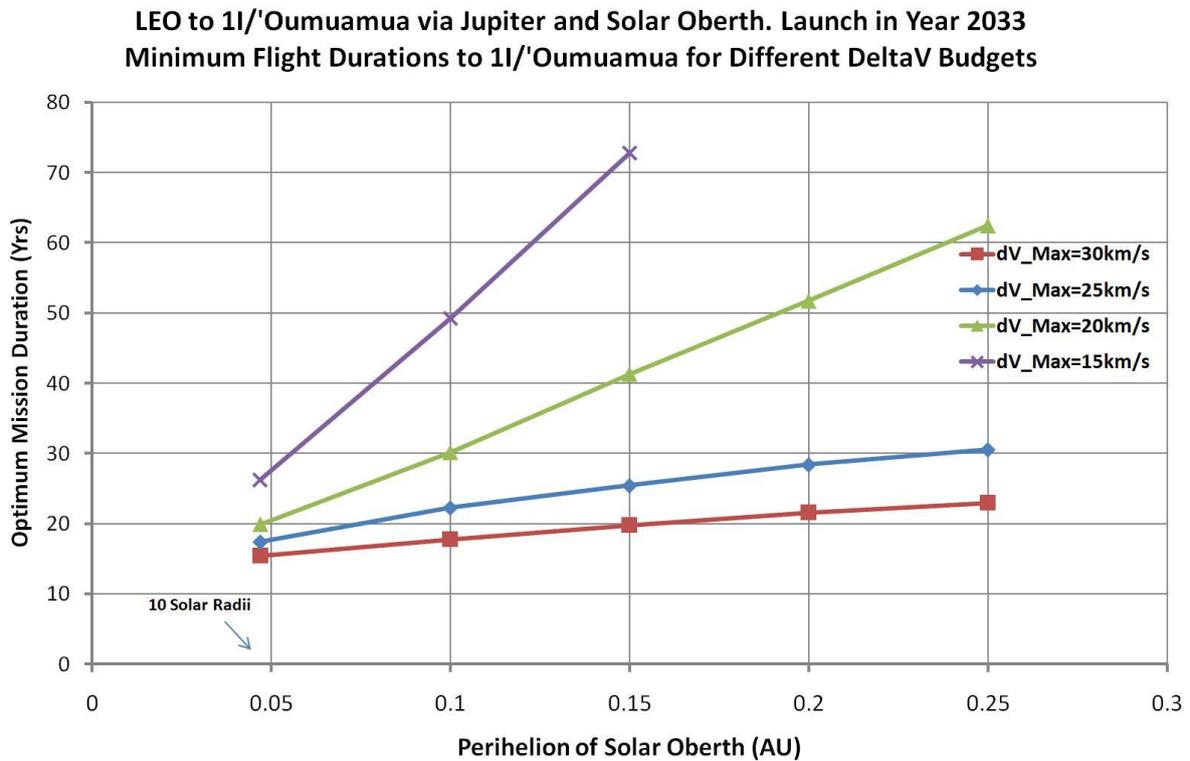

*Figure 8 : Scenario (3) Minimum Mission Durations vs Solar Oberth Distance for 4 ΔV budgets*

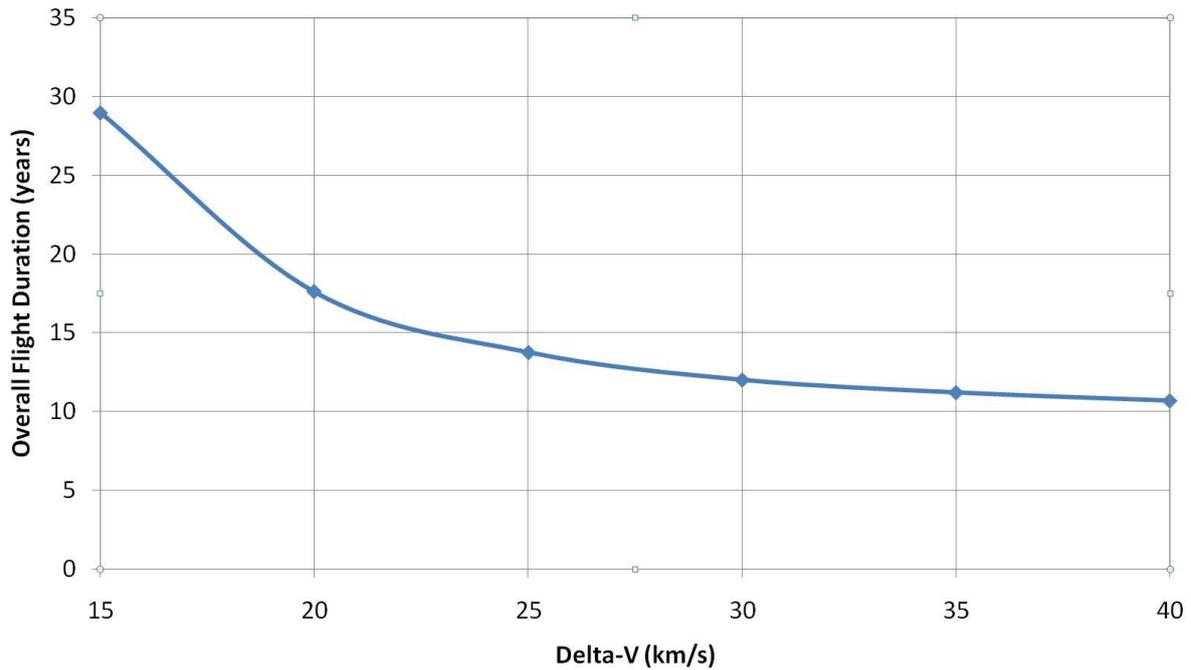

*Figure 9 : Scenarios (4) & (5) Mission Duration vs ΔV budget*

## 3.2 Payload Masses to 'Oumuamua Achievable Using NTR

For direct trajectories, i.e. scenario (1), data from Figure (6) can be combined with Figure (1) to give Figure (10).

Data from Figures (7), (8) and (9) can be combined with Figure (1) to give Figures (11)-(13). Figure (11) is scenario (2) and note that the NERVA Pewee-class engine is excluded from this plot because it cannot achieve payload masses > 0kg since the required ΔV budgets are too high.

The results are summarised in Table 2.

Regarding these Figures (10)-(13), it can be observed that there is a clear relationship between payload mass and flight time. Hence as one might expect, as flight time goes up so the ΔV reduces enabling higher payload masses (ref. Figure (1)). Generally of the NTR options, the NERVA Pewee-class NTR has the worst performance (in terms of longer flight times and lower payload masses) whereas the Particle Bed (SNTP) has the best performance. As also might be predicted, for those trajectories which employ a Solar Oberth Maneuver, the closer the Solar Oberth to the sun, the better the overall mission capability, though naturally the solar flux and consequent heat shield mass requirement increase.

Scenario (2) enables lower flight durations than scenario (1) but has lower payload mass capability.

Generally scenario (3) enables higher payload masses to 1I than scenarios (1) & (2) but it can also be seen that scenarios (1) & (2) trajectories have yearly optima as opposed to the 12-yearly optima for scenario (3). Furthermore, flight durations are lower for scenarios (1) & (2). Scenario (4) with $LH_2$ tanks is shown in Figure (13) and offers better performance than scenario (3), but without a hazardous close approach to the sun. Scenario (5) with $NH_3$ tanks is shown in Figure (14) and has lower payload masses compared to $LH_2$ as would be expected.

Comparing Figure (13) (scenario (4) using a powered Jupiter GA with $LH_2$ tanks) against Figure (10) (scenario (1), direct transfer), it appears that the former provides significantly higher payload masses, but this is only because the former has a generally lower ΔV mission profile. In fact for equivalent ΔV budgets, these two scenarios have the same payload mass but scenario (4) gives a significantly lower flight duration. So for instance, with ΔV= 25km/s, scenario (1) gives a duration of 37 years whereas scenario (4) with $LH_2$ has a duration of 14 years. However there are two key disadvantages with scenario (4) and they are first that it involves a journey to Jupiter which requires the $LH_2$ to be stored without significant leakage and with zero boil-off (so with a cryocooler) and second the launch optima are at twelve year intervals, between which trajectories are not viable.

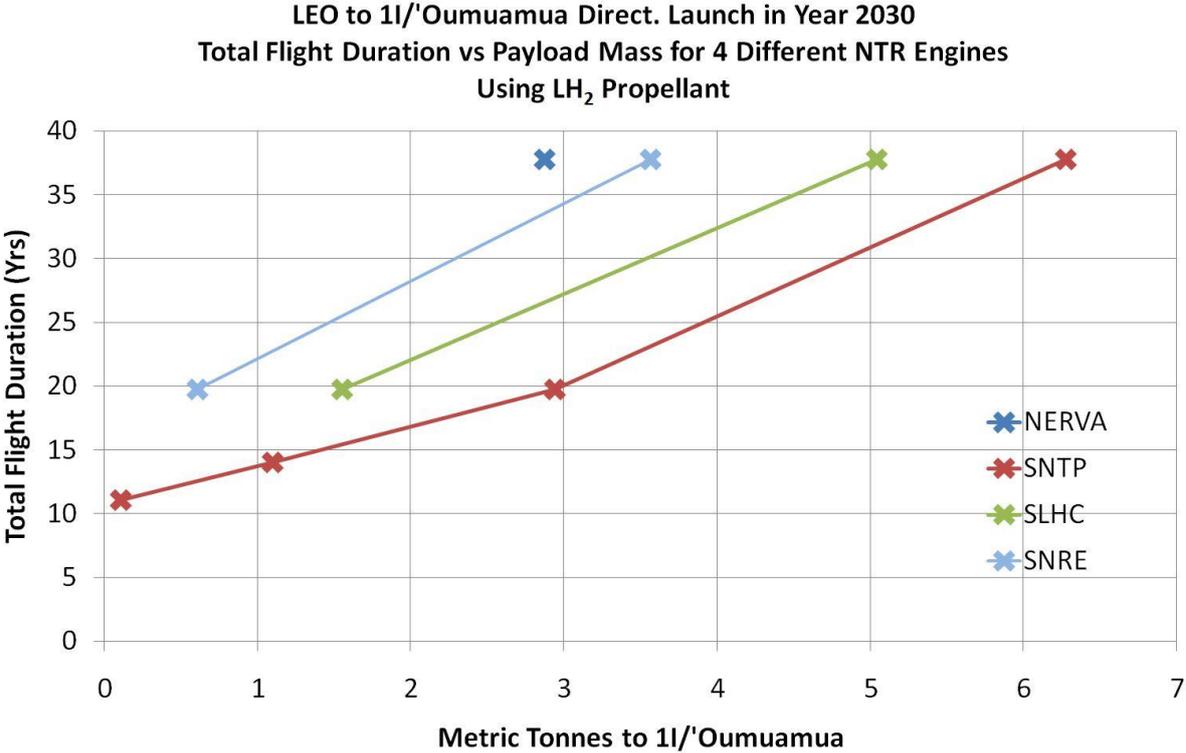

Figure 10 : Scenario (1) Flight Duration vs Mass to 1I/'Oumuamua

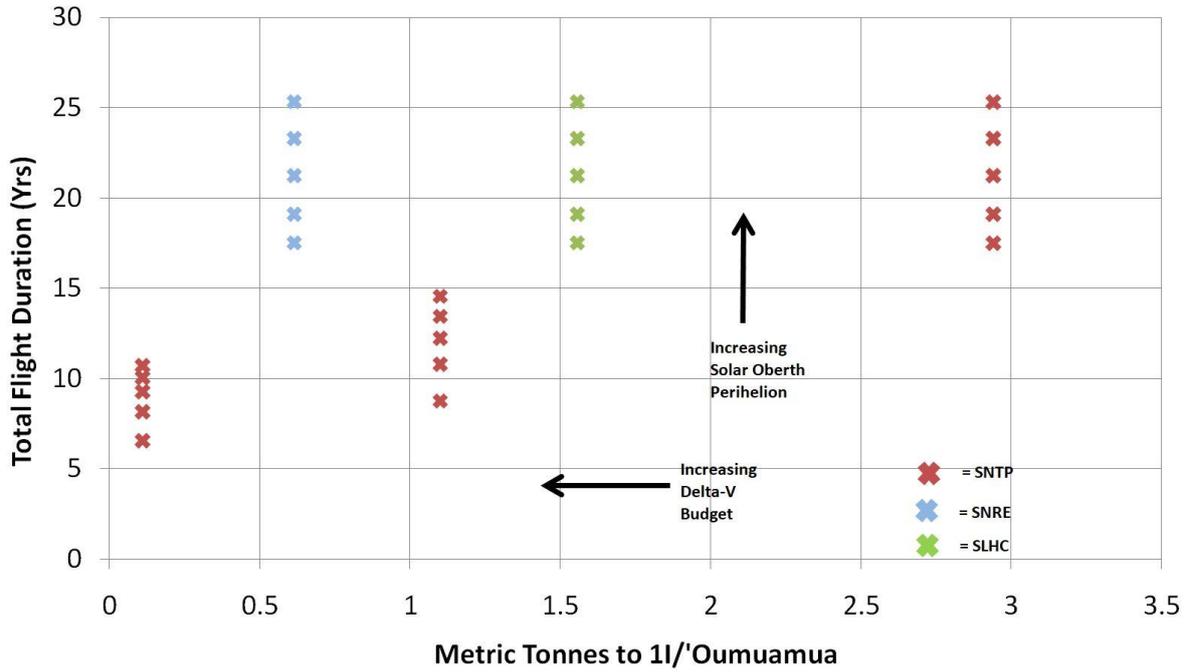

*Figure 11 : Scenario (2) Flight Duration vs Mass to 'Oumuamua*

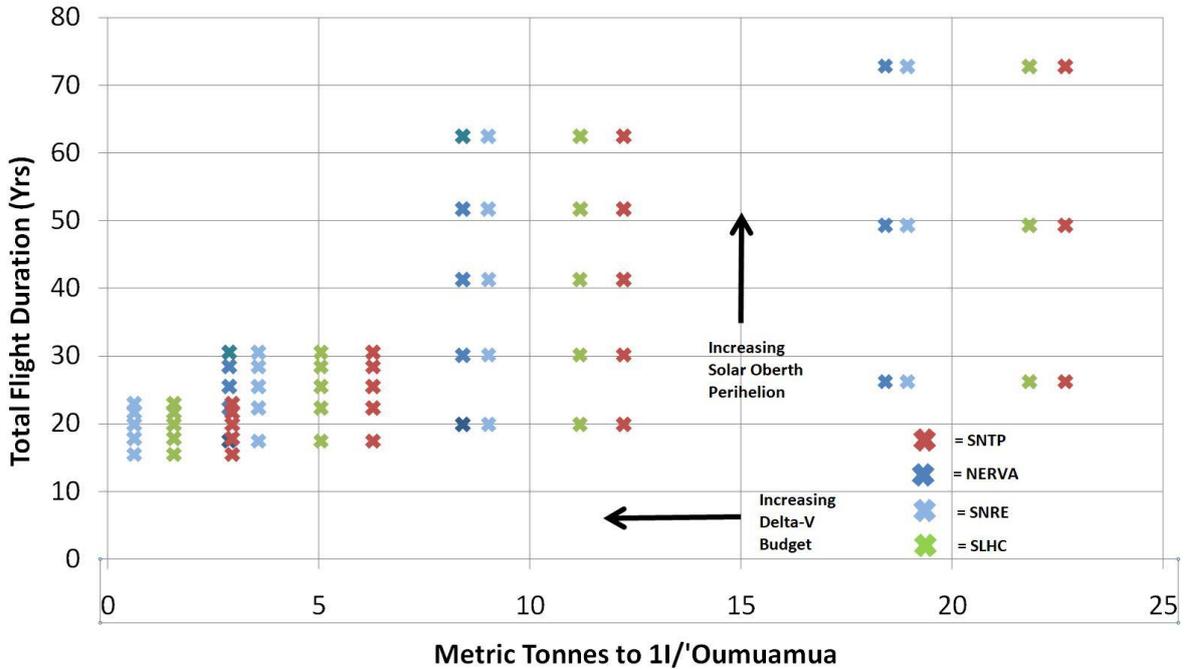

*Figure 12 : Scenario (3) Flight Duration vs Mass to 'Oumuamua*

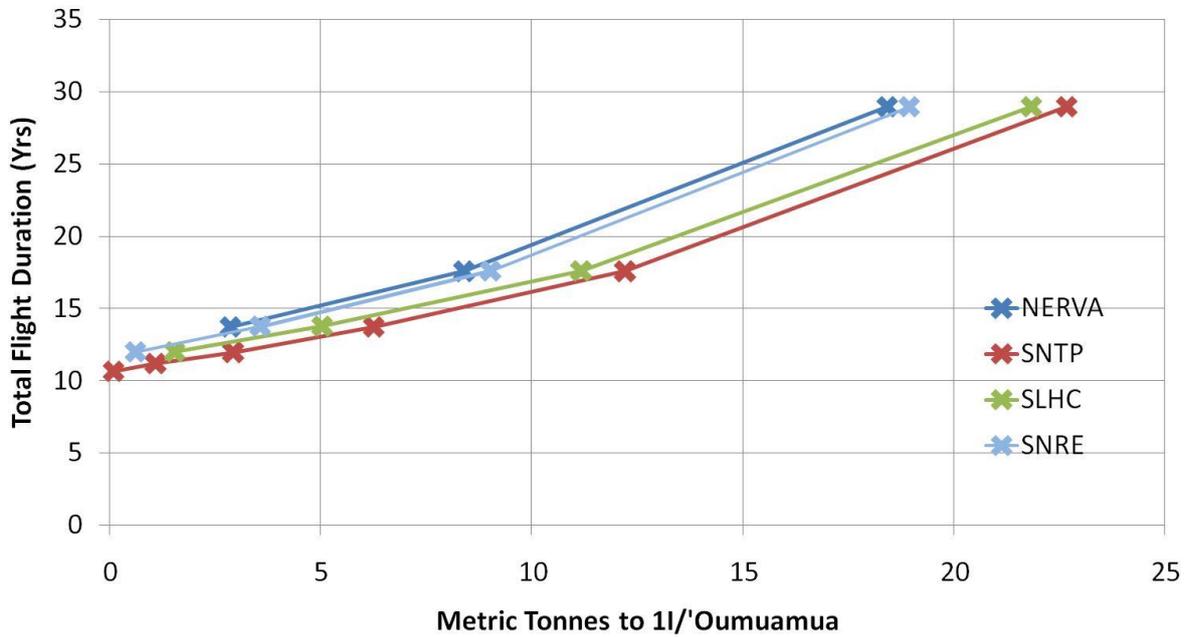

*Figure 13 : Scenario (4) with Liquid Hydrogen, LH2 propellant.*

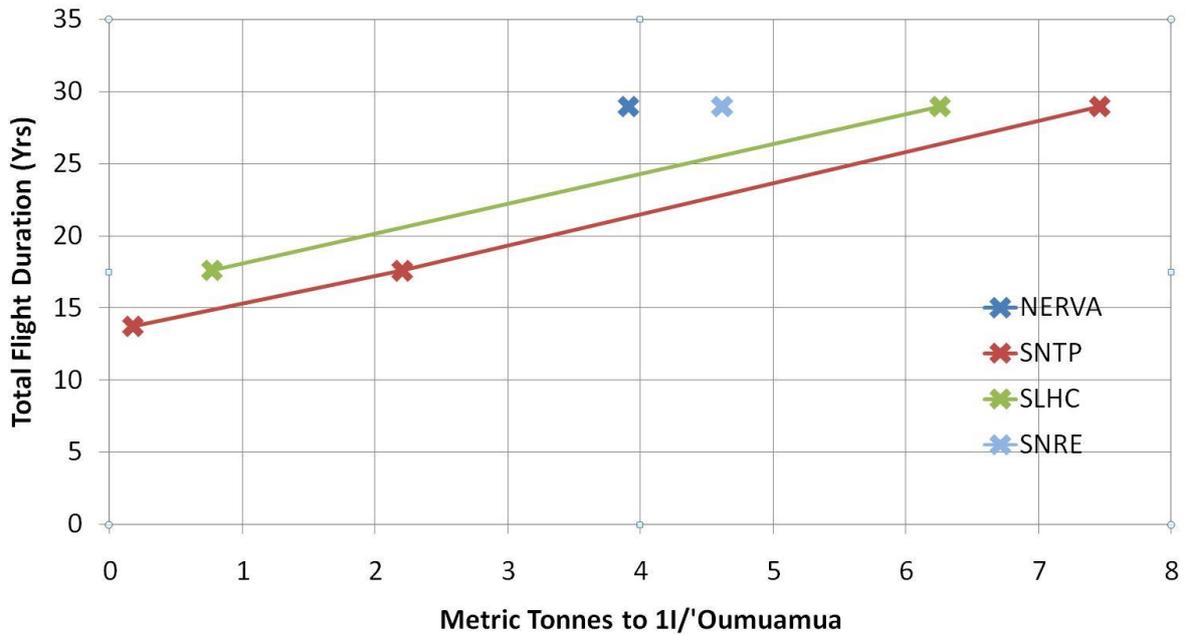

*Figure 14 : Scenario (5) with Ammonia, $NH_3$ propellant*

Table 2 Results Summary

| Trajectory Scenario | Trajectory description | Optimum Launch Windows | Mass achievable using Pewee-class NTR/mt | Mass achievable with SNTP/mt | Average mass achievable/ mt | Minimum Flight Duration/yrs | Maximum Flight Duration/yrs | Average Flight Duration/yrs | Relative Arrival Velocity (km/s) for $\Delta V=30$km/s |
|---|---|---|---|---|---|---|---|---|---|
| 1 | Direct from LEO to 1I | Every year | 2.9 ($\Delta V=25$km/s) | 6.3 ($\Delta V=25$km/s) | 2.7 | 11.1 | 38.0 | 26.2 | 17.8 |
| 2 | LEO to Solar Oberth to 1I | Every year | 0 | 2.9 ($\Delta V=30$km/s) | 1.3 | 6.5 | 25.3 | 21 | 19.4 |
| 3 | LEO to Jupiter to Solar Oberth to 1I | Every 12 years, 2033, 2045, 2057,…. | 18.4 ($\Delta V=15$km/s) | 22.7 ($\Delta V=15$km/s) | 7.8 | 15.2 | 72.8 | 32 | 43.1 |
| 4 | LEO to Jupiter to 1I using LH$_2$ | Every 12 years, 2031, 2043, 2055,…. | 18.4 ($\Delta V=15$km/s) | 22.7 ($\Delta V=15$km/s) | 8.6 | 10.7 | 29.0 | 17.6 | 39.7 |
| 5 | LEO to Jupiter to 1I using NH$_3$ | Every 12 years, 2031, 2043, 2055,…. | 3.9 ($\Delta V=15$km/s) | 7.5 ($\Delta V=15$km/s) | 3.6 | 13.8 | 29.0 | 23.6 | 39.7 |

# 4 Discussion

In this paper, we investigated the use of NTR for chasing interstellar objects, once they have left the inner solar system. We used the example of 1I/'Oumuamua for illustration.

We identified several advantages of using NTR for a mission to 'Oumuamua and similar interstellar objects. First, due to the higher Isp, flight durations can be considerably reduced to < 15 years compared to > 20 years for chemical propulsion. The higher Isp also implies that generally the payload masses to 1I/'Oumaumua are considerably larger than for chemical propulsion. This translates to payload masses of 1000's of kg as opposed to 100's of kg for chemical.

In terms of the trajectories, direct trajectories are possible, which significantly reduce the complexity of missions. Direct trajectories also mean that the optimum launch windows arrive once a year, when the Earth and 'Oumuamua are appropriately aligned. Second, in contrast to chemical propulsion, the arrival velocities are much lower approx. 18km/s, compared to 30km/s with chemical. Lower velocities allow for longer observation times during the encounter and thereby a higher science return. Trajectories without a Solar Oberth maneuver also have the advantage of a lower degree of uncertainty. One of the caveats of the Solar Oberth maneuver is that the errors or uncertainties in the burn at Perihelion have a disproportionate influence on the solar system escape trajectory. Also, the heat shield is not required, thereby saving mass. If a Solar Oberth is utilized, they can be farther from the sun (perihelia for the SO can be > 10Solar Radii) compared to chemical propulsion (where perihelia for the SO are typically< 10 Solar Radii), which reduce the requirements for the heat shield, as the solar irradiation per area diminishes with $1/r^2$.

A major drawback of the NTR trajectories which employ a trip to Jupiter, is the high relative velocity of the s/c as it approaches 1I/'Oumuamua (from Table 2 around 40km/s).

However, though not studied, there is the potential for slowing down as the target is approached and even to perform a rendezvous, though this would be after a long flight duration and so contingent on nearly zero-leakage and zero boil-off $LH_2$ tanks and cryocoolers.

The direct trajectory option (scenario (1)) would not require long $LH_2$ storage durations and for reasons elucidated above is possibly one of the preferred options. However if mission duration is important, scenario (4), Earth-Jupiter-1I, is possibly the preferred choice.

One limitation of our study is that the payload is considered as a black box, and potential constraints from the spacecraft with its instrumentation are not taken into consideration. Such constraints might be related to compatibility issues between NTR and certain spacecraft instruments.

To summarize, our findings indicate that NTR for missions to interstellar objects would have a significant effect on the duration of a mission to such objects (trip times can be even cut in half) and allow for payload masses of an order of magnitude higher than for chemical propulsion. Hence, NTR would be a game changer for missions chasing interstellar objects, when they are on their way out of the solar system. Future work should explore the use of NTR for rendezvous and even sample return missions, which are feasible with NTR.

# 5 Conclusions

In this paper, we examined the use of Nuclear Thermal Propulsion (NTP) for missions to interstellar objects, exemplified by 1I/'Oumuamua. Four different proposed NTP options are analysed, ranging from NERVA-based designs to more advanced NTP. Using the OITS trajectory optimization tool, we find that NTP would allow for simpler and more direct, time-saving trajectories to 1I/'Oumuamua. Significant savings in terms of mission duration (14 years for a launch in 2031) are identified. Payload masses on the order of 1000s of kg, compared to 100s of kg using a Space Launch System launcher would be feasible. We conclude that NTP would be a game changer for chasing interstellar objects on their way out of the solar system, drastically reducing trip times and increasing payload masses. Future work should explore rendezvous missions using NTP as well as the feasibility of using NTP for reactive missions, where interstellar objects are discovered early.